\documentclass[aps,prd,twocolumn,amsmath,amssymb,superscriptaddress,nofootinbib,floatfix,showpacs]{revtex4-1}

\usepackage{graphicx}
\usepackage{color}
\usepackage{subfigure}
\usepackage{xspace}

\newcommand{\dslash}{\ensuremath{\partial\hspace{-1.2ex} /}}
\newcommand{\Dslash}{\ensuremath{D\hspace{-1.5ex} /}}
\newcommand{\Tr}{\ensuremath{\operatorname{Tr}}}

\def\lsim{\raise0.3ex\hbox{$<$\kern-0.75em\raise-1.1ex\hbox{$\sim$}}}
\def\gsim{\raise0.3ex\hbox{$>$\kern-0.75em\raise-1.1ex\hbox{$\sim$}}}

\def\roughly#1{\mathrel{\raise.3ex\hbox{$#1$\kern-.75em%
\lower1ex\hbox{$\sim$}}}}

\newcommand{\vev}[1]{\ensuremath{\left\langle #1 \right\rangle}}
\newcommand{\einh}[1]{\ensuremath{\,\text{#1}}}
\newcommand{\MeV}{\einh{MeV}}

\def\Eq#1{Eq.~(\ref{#1})}
\def\Fig#1{Fig.~\ref{#1}}

\newcommand{\ua}{\ensuremath{U(1)_A}}
\newcommand{\Phibar}{\ensuremath{\bar{\Phi}}}
\newcommand{\LPQM}{\ensuremath{\mathcal{L}_{\textrm{PQM}}}\xspace}

\newcommand{\muT}{\left(\frac{\mu}{T}\right)}
\newcommand{\onefig}{0.9\linewidth}

\newcommand{\threefigs}{0.3\linewidth}
\newcommand{\pade}{Pad\'e}

\graphicspath{
{./}
{../figures/}
{./figures/}
}

\begin{document}

\title{Towards finite density QCD with Taylor expansions}
 \author{F.~Karsch}
\email[E-Mail:]{karsch@physik.uni-bielefeld.de, karsch@bnl.gov} \affiliation{Physics
  Department, Brookhaven National Laboratory,Upton, NY 11973, USA}
\affiliation{Fakult\"at f\"ur Physik, Universit\"at Bielefeld, D-33615
  Bielefeld, Germany} \author{B.-J.~Schaefer}
\email[E-Mail:]{bernd-jochen.schaefer@uni-graz.at}
\affiliation{Institut f\"{u}r Physik, Karl-Franzens-Universit\"{a}t,
  A-8010 Graz, Austria} \author{M.~Wagner}
\email[E-Mail:]{mathias.wagner@physik.tu-darmstadt.de}
\affiliation{Institut f\"{u}r Kernphysik, TU Darmstadt, D-64289
  Darmstadt, Germany} \affiliation{ExtreMe Matter Institute EMMI, GSI
  Helmholtzzentrum f\"ur Schwerionenforschung mbH, Planckstr. 1,
  D-64291 Darmstadt, Germany} \author{J.~Wambach}
\email[E-Mail:]{jochen.wambach@physik.tu-darmstadt.de}
\affiliation{Institut f\"{u}r Kernphysik, TU Darmstadt, D-64289
  Darmstadt, Germany} \affiliation{Gesellschaft f\"{u}r
  Schwerionenforschung GSI, D-64291 Darmstadt, Germany.} \date{\today}


\begin{abstract}
  Convergence properties of Taylor expansions of observables, which
  are also used in lattice QCD calculations at non-zero chemical
  potential, are analyzed in an effective $N_f = 2+1$ flavor
  Polyakov-quark-meson model. A recently developed algorithmic
  technique allows the calculation of higher-order Taylor expansion
  coefficients in functional approaches. This novel technique is for
  the first time applied to an effective $N_f = 2+1$ flavor
  Polyakov-quark-meson model and the findings are compared with the
  full model solution at finite densities. The results are used to
  discuss prospects for locating the QCD phase boundary and a possible
  critical endpoint in the phase diagram.
\end{abstract}

\maketitle

\section{Introduction}

Exploring chiral and deconfining properties of strongly-interacting
matter at high temperature is one of the most interesting issues in
modern physics (for reviews, see e.g.~
\cite{Kogut:2004su, Fukushima:2010bq, Rajagopal:2000wf}).
In particular, the possible existence of a critical endpoint (CEP) in
the QCD phase diagram at non-zero net baryon number density and its
consequences for the phase structure of QCD at lower temperatures are
much under debate, e.g.~\cite{Stephanov:2004wx}. High energy heavy-ion
collision experiments at RHIC and SPS started to look for experimental
evidence of the CEP and experiments at new heavy-ion facilities (FAIR
and NICA), have been designed to probe the relevant high density
region in the QCD phase diagram.

Gaining insight into properties of QCD at non-zero, even small values
of the quark chemical potential ($\mu$) is thus of great interest.
However, in theoretical calculations the relevant region in the QCD
phase diagram is not easy to access directly in QCD. Model
calculations or non-perturbative lattice QCD calculations, based on
approximation schemes, are currently used.

At vanishing chemical potential the lattice QCD calculations get
refined systematically and become possible with almost realistic quark
mass values. They indicate a temperature-driven crossover transition
from hadrons to the quark-gluon-plasma phase, for a recent review see~\cite{DeTar:2009ef}.

However, even in this case the quantitative determination of the
transition temperature is still a difficult task. A careful analysis
of cut-off effects is needed to extract this quantity in the continuum
limit~\cite{Cheng:2009be, *Bazavov:2009zn, *Cheng:2006qk,
  Borsanyi:2010bp, *Aoki:2009sc, *Aoki:2006br}.

At non-zero values of the chemical potential the notorious fermion
sign problem prohibits straightforward lattice QCD simulations. In
order to enter at least the small chemical potential region of the QCD
phase diagram several extrapolation methods such as the reweighting
technique, analytic continuation to imaginary chemical potential or
Taylor series expansions have been proposed. All these extrapolation
methods have their own, intrinsic problems and their reliability is
still under investigation, see e.g.~\cite{Philipsen:2005mj,Schmidt:2006us, deForcrand:2010ys}
for reviews. Under these circumstances, it may be
useful to apply approximation schemes, used currently in lattice QCD
calculations, also in model calculations. Through a comparison with
analytic results obtainable in QCD-like models, insight may be gained
for the application of approximation schemes in lattice QCD
calculations.

In this work we focus on the Taylor expansion technique where a
thermodynamic quantity such as the pressure is expanded in powers of
the chemical potential $\mu$ around vanishing chemical potential,
i.e., around the point at which Monte Carlo simulations are not
hindered by the sign problem. This method was developed for studies of
QCD thermodynamics with the aim to gain information also on the
location of a possible critical point in the QCD phase diagram
\cite{Allton:2002zi, *Allton:2003vx, *Allton2005gk, Gavai:2003mf, *Gavai:2004sd,*Gavai:2008zr}. In
general, the relevant Taylor coefficients can be calculated with
standard simulation techniques used also at $\mu =0$
\cite{Bernard:1996cs, Karsch:2000ps, AliKhan:2001ek}. With this
approach the calculation of several thermodynamic quantities has been
extended to non-zero values of the chemical potential; results have
been compared with hadron resonance gas model calculations at low
temperature~\cite{Karsch:2003vd, *Karsch:2003zq} and quark-gluon gas
thermodynamics at high temperature. Attempts have also been made to
determine the phase boundary and to infer a possible critical endpoint
in the phase diagram from convergence properties of the Taylor series
for thermodynamic quantities, e.g., the pressure or quark number
susceptibility~\cite{Allton:2002zi, *Allton:2003vx, *Allton2005gk, Gavai:2003mf, *Gavai:2004sd,*Gavai:2008zr, Schmidt:2006us, Schmidt:2007jg, *Miao:2008sz}.

So far, basic features of the Taylor series such as its convergence
properties have received only little attention because a proper
investigation of these aspects requires the knowledge of higher-order
Taylor coefficients. In most QCD studies these coefficients are
difficult to obtain and a systematic analysis of convergence
properties is not yet possible. At present any attempts to identify
some precursor effects of a possible critical endpoint thus have to
rely on information extracted from a few low order expansion
coefficients.

Recently, a new algorithmic differentiation technique has been
developed that allows for the calculation of higher-order Taylor
coefficients in functional approaches \cite{Wagner:2009pm}. This
enables one to investigate general convergence properties of the
Taylor expansion method at least in model calculations. Their
consequences and prospects for the determination of the phase boundary
and a possible critical endpoint are discussed here in the framework
of the effective $N_f = 2+1$ flavor Polyakov--quark-meson (PQM) model.
The model is studied in mean-field approximation and the findings from
the Taylor expansion are compared with the corresponding model
mean-field solution at finite density~\cite{Schaefer:2009st, *Wambach:2009ee}.

The outline of this work is as follows: In Sec.~\ref{sec:taylor} the
Taylor expansion method and its application to the analysis of
different approximants for radii of convergence are introduced. We
also discuss the improvement of Taylor series expansions by a Pad\'e
resummation. Basic strategies for the determination of the phase
boundary and the location of a critical endpoint in the phase diagram
from Taylor expansion coefficients are discussed in
Sec.~\ref{sec:locating}. Sec.~\ref{sec:model} is devoted to the
presentation of the model analysis where higher-order Taylor
coefficients are used to locate the phase boundary and the critical
endpoint at finite chemical potential. We conclude in
Sec.~\ref{sec:summary} with a brief summary and discussion of our
findings.

\section{Extrapolation to finite $\mu$}
\label{sec:taylor}

The Taylor expansion method has been used as an approximation scheme
to gain information on QCD thermodynamics at high temperature and for
small, non-vanishing values of the chemical potential. Here, we start
with a description of some generic model-independent properties of the
Taylor expansion of, e.g., the QCD partition function (see also
\cite{Schmidt:2009qq}).

\subsection{Taylor expansion}

At fixed temperature the logarithm of the QCD partition function, {\it i.e.}
the pressure $p = (T/V) \ln \mathcal{Z}$, may be expanded in a
Taylor series in powers of $\mu/T$ around~\mbox{$\mu=0$:}
\begin{equation}
  \label{eq:pressure}
  \frac{p(\mu/T)}{T^4} = \sum_{n=0}^\infty c_n(T) \muT^n\ .
\end{equation}
The Taylor expansion coefficients are determined by derivatives
with respect to the chemical potential evaluated at vanishing $\mu$,
\begin{equation}
  \label{eq:pcoeff}
  c_n(T) = \left.\frac{1}{n!} \frac{\partial^n\left(p(T,\mu) /T^4
      \right)}{\partial \left(\mu/T\right)^n} \right|_{\mu=0}\ . 
\end{equation}
Due to the CP-symmetry of QCD, the partition function is symmetric in
$\mu$, {\it i.e.} $\mathcal{Z}(\mu)=\mathcal{Z}(-\mu)$ which reflects
the invariance of $\mathcal{Z}$ under exchange of particles and
anti-particles. As a consequence, all odd expansion coefficients
vanish and the series is even in $\mu/T$. In QCD, various quark
chemical potentials have to be taken into account in this expansion.
However, in order to simplify our discussion in the following, we
focus on one uniform quark chemical potential $\mu \equiv \mu_q =
\mu_B/3$ which will not restrict our conclusions.

Besides the pressure, any thermodynamic observables like, e.g., the
energy density ($\epsilon$), the trace anomaly $(\epsilon-3p)$ or the
quark number susceptibility $\chi_q$ can be expanded in a similar
series with appropriately modified Taylor expansion coefficients
\cite{Miao:2008sz, DeTar:2010xm}. Of course, all these modified Taylor
expansion coefficients are related to the initial expansion
coefficients of the pressure. An instructive example is the Taylor
expansion of the quark number susceptibility, $\chi_q$, which is
defined as the second derivative of the pressure w.r.t. to $\mu$
\begin{equation}
  \label{eq:suscep}
  \chi_q(T,\mu) = \frac{\partial^2 p(T,\mu)}{\partial \mu^2}\ .
\end{equation}
This yields the following relation to the corresponding Taylor
expansion coefficients $c_n (T)$ of the pressure
\begin{equation}
  \label{eq:taylorsuscep}
  \frac{\chi_q(\mu/T)}{T^{2}} = \sum_{n=2}^{\infty}
  \underbrace{n (n-1) c_n(T)}_{ =c^\chi_{n-2}(T)}
  \left(\frac{\mu}{T}\right)^{n-2}\ 
  .
\end{equation}
The expansion of other thermodynamic quantities like, for instance,
the trace anomaly requires additional derivatives of the Taylor
coefficients w.r.t. the temperature.

In the following we compare the Taylor expansion of the pressure and
the quark number susceptibility at non-zero chemical potential. If a
critical endpoint exists in the QCD phase diagram then it is expected
that the quark number susceptibility diverges at that point while the
pressure stays finite and is still continuous. Although the Taylor
series for both quantities have identical radii of convergence they
thus will behave differently on the phase boundary. A comparison of
both series as the truncation order is increased might therefore be
helpful to gain new insights in locating a possible CEP in the QCD
phase diagram.

\subsection{Radius of convergence}

Whether or not the radius of convergence of the Taylor expansion of
the thermodynamic potential around $\mu=0$ is related to a phase
transition in the physical system crucially depends on the location of
the singularity in the complex $\mu$-plane that causes the breakdown
of the series expansion. Only if this singularity lies on the real
axis the radius of convergence corresponds to the location of a
critical point. Information on the location of the singularity in the
complex $\mu$-plane can be deduced from properties of the Taylor
expansion coefficients themselves. In any case, the singularity
closest to $\mu=0$ will determine the radius of convergence of the
Taylor series. This can be obtained from the behavior of the Taylor
expansion coefficients of the pressure,
\begin{equation}
  r = \lim_{n \rightarrow \infty} r_{2n} = \lim_{n \rightarrow \infty}
  \left|\frac{c_{2n}}{c_{2n+2}} \right|^{1/2}\ . 
\label{eq:convr}
\end{equation}
Other definitions of the radius of convergence are known and their
application in the context of QCD have been discussed in the
literature, see e.g.~\cite{Gavai:2003mf, *Gavai:2004sd, *Gavai:2008zr}.

Of course, in almost all applications only a finite number of
expansion coefficients are known and the limit $n\rightarrow\infty$
cannot be performed. In some cases it might be possible to extrapolate
the $r_{2n}$ to infinity, making use of their expected asymptotic
behavior. Otherwise, one will have to assume that the radius of
convergence $r$ has been reached when subsequent estimators, $r_{2n}$, do
not change within errors anymore.

At a given order $n$ the Taylor expansions of different observables will
give different estimators for the radius of convergence, which are all
related. For example, the estimator for the radius of convergence
obtained from the quark number susceptibility, $r_{2n}^\chi$, is
related to the one obtained from the pressure series via
\begin{equation}
  \label{eq:convradchi}
  r_{2n}^\chi = \left|\frac{c^\chi_{2n}}{c^\chi_{2n+2}} \right|^{1/2}=
  \left(\frac{(2n+2)(2n+1)}{(2n+3)(2n+4)}\right)^{1/2}r_{2n+2} \ .
\end{equation}
In the limit $n \rightarrow \infty$ both estimators for the radius of 
convergence will converge to the same unique limit $r$ since the 
prefactor in front of $r_{2n+2}$ tends to one. However, at finite order $n$
the estimates for $r_{2n}$ and $r_{2n}^\chi$ will differ and as a consequence 
both may approach the limit $r$ at a different rate. 
For large $n$ we find from \Eq{eq:convradchi}
\begin{equation}
  \label{eq:convradchi2}
 r_{2n}^\chi = r_{2n+2} \left( 1 - \frac{1}{n} + {\cal O}(n^{-2})\right) \; .
\end{equation}
When deviations from the asymptotic value $r$ are of ${\cal O}(1/n)$,
i.e., $r_{2n+2} \simeq r (1+A/n)$ with a positive, constant $A$, both estimators
for the radius of convergence will converge with the same truncation
errors of ${\cal O}(1/n)$ (cf. App.~\ref{sec:crit}). Deviations from the asymptotic value $r$
will, however, be numerically smaller in the susceptibility series
than in the pressure series by a factor $(A-1)/A$. In other words, the
estimator for the radius of convergence obtained from the
susceptibility series at given order $n$ might be more suitable for
estimating the true radius of convergence than the one obtained from
other observables, e.g., from the pressure series.

\subsection{Resummations: Pad\'e approximation}
\label{sec:pade_approx}

The convergence of a Taylor series can be improved further by a
resummation of the expansion coefficients which is based on a Pad\'e
approximation (for details see \cite{Baker1965, *Baker:312952}). For
clarity and completeness, we collect the needed facts here. The Pad\'e
resummation employs the same derivative information as the initial
Taylor expansion coefficients but often shows an extended convergence
range, in particular in the presence of singularities. Furthermore, a
more stable result can be achieved with fewer coefficients.

The Pad\'e approximant can be obtained by rewriting a Taylor series of
an analytic function $t(x)$ around $x_0=0$ to order $N$
\begin{equation}
  t(x)=\sum\limits_{i=0}^N t_{(i)} x^{i}
\end{equation}
as a ratio of two power series $p(x)$ and $q(x)$ of order $L$ and $M$,
respectively
\begin{equation}
  [L/M]\equiv
  R_{L,M}(x)=\frac{p(x)}{q(x)}=\frac{p_0+p_1x+\cdots+p_Lx^L}{1+q_1
    x+\cdots+q_Mx^M}\ . 
\end{equation}

One advantage of this reformulation is that poles and singularities
are better characterized by a rational function than by a polynomial
Taylor series. The expansion coefficients for $p$ and $q$ can be
obtained up to order $N=L+M$ by equating the derivatives of both
series up to order $N$ at $x=0$, i.e.,
\begin{equation}
  \left.\frac{\partial^i R_{L,M}(x)}{\partial x^i}\right\vert_{x=0} =
  \left.\frac{\partial^i t(x)}{\partial x^i}\right\vert_{x=0} \quad
  \text{for } i \leq N\ . 
\end{equation}

An alternative and more elegant form relates the Pad\'e approximant to
determinants of two $(M+1)\times (M+1)$ matrices as follows
\begin{equation}
\label{eq:pade}
[L/M]= \small \frac{\left| \begin{array}{cccc}
  t_{(L-M+1)} & t_{(L-M+2)} & \cdots & t_{(L+1)}\\
  t_{(L-M+2)} & t_{(L-M+3)} & \cdots & t_{(L+2)}\\
  \vdots & \vdots & \ddots & \vdots\\
  t_{(L)} & t_{(L+1)} & \cdots & t_{(L+M)}\\
  \sum\limits_{i=0}^{L-M} t_{(i)} x^{M+i} & \sum\limits_{i=0}^{L-M+1}
  t_{(i)}x^{M+i-1} & \cdots & \sum\limits_{i=0}^{L} t_{(i)}x^{i} 
\end{array}
\right|}
{\left|
\begin{array}{ccccc}
  t_{(L-M+1)} & t_{(L-M+2)} & \cdots & t_{(L+1)}\\
  t_{(L-M+2)} & t_{(L-M+3)} & \cdots & t_{(L+2)}\\
  \vdots & \vdots & \ddots & \vdots\\
  t_{(L)} & t_{(L+1)} & \cdots &  t_{(L+M)}\\
  x^{M} & x^{M-1} & \cdots & 1
\end{array}
\right|}\,.
\end{equation}
Note that for $M=0$ and $L=N$ the Pad\'e approximant is identical to
the initial Taylor approximation, i.e., $[N/0] \equiv t(x)$. Of
course, apart from choosing the highest derivative order $N$ in the
\pade{} approximation one can also vary $L$ or $M$ independently.

The determination of the radius of convergence or the phase boundary
with the \pade{} approximation is more involved. For the case $[N/2]$
one immediately sees, \Eq{eq:pade}, that the \pade{} approximant has a
pole at $x=\pm \sqrt{c_{N}/c_{N+2}}$ since the odd Taylor coefficients
vanish. This pole coincides with the estimator for the radius of
convergence, $r_{N}$, of the Taylor series up to order $x^{N+2}$,
\Eq{eq:convr}. For a general \pade{} approximant $[L/M]$ the pole
structure is more elaborated. However, in this case the first pole at
positive and real chemical potential might provide a useful estimate
for the phase boundary.

\section{Locating a critical endpoint}
\label{sec:locating}

It has been argued that the estimators for the radius of convergence
can be used to determine the location of a possible CEP in the QCD
phase diagram \cite{Allton:2003vx, Gavai:2004sd, Stephanov:2006dn,
  Gavai:2008zr}. In general, a breakdown of a Taylor expansion is
caused by a singularity in the complex $\mu$-plane. Only if this
singularity, which is closest to the origin at $\mu=0$, lies on the
real $\mu$-axis, it is also related to a physical phase transition
since it is a singularity in a thermodynamic quantity corresponding to
a zero in the partition function \cite{Yang:1952be}. If a CEP exists
in the QCD phase diagram, it belongs to the universality class of the
three dimensional Ising model. The corresponding second-order phase
transition will lead to divergences in thermodynamic observables with
a power law behavior.

The finite radius of convergence arises from a singularity on the real
$\mu$-axis only if there exists a $n_0$ such that for all $n>n_0$ all
expansion coefficients are positive. Since in practice only a few
terms in the Taylor series are known such a mathematically rigorous
conclusion cannot be drawn in general. The hope is that by
investigating the structure of a few known coefficients some hints on
the properties of the Taylor series and an estimate for the location
of a CEP in the phase diagram can be obtained. Through an inspection
of the relative magnitude and relative signs of the expansion
coefficients the determination of the radius of convergence of the
Taylor series might be possible and the singularity in the complex
$\mu$-plane which causes the breakdown of this expansion can be
located.

Since the thermodynamics in the vicinity of a second-order phase
transition is dominated by the singular contribution of the free
energy density or thermodynamic potential a generic structure of the
Taylor expansion coefficients should occur \cite{Karsch:2009zz}. At
vanishing chemical potential the singular behavior near a phase
transition is controlled by the reduced temperature which depends
linearly on $T-T_c$ but quadratically on the chemical potential. Thus,
a $(2n)^{th}$ derivative with respect to $\mu$ shows a singular
behavior that is similar to that of the $n^{th}$ derivative with
respect to temperature. This infers that the structure of the coefficient
$c_{2n+2}$ can be estimated by investigating the temperature
derivative of $c_{2n}$. This is indeed found in our model
calculations. For an example see Fig.~\ref{fig:coeff} and the
discussion in Sec.~\ref{sec:ad}. This means that $c_{2n+2}$ will stay
positive only up to a temperature which corresponds to the first
maximum of the previous coefficient $c_{2n}$. Near the transition
temperature the coefficients start to oscillate. Thus, through a
determination of the first maximum in $c_{2n}$, an estimate for the
largest temperature below which all coefficients are
positive\footnote{As pointed out earlier the mathematically more
  rigorous criterion for a singularity on the real axis does not rule
  out the possibility to have a few negative expansion coefficients.
  We assume here that once a negative coefficient occurs this will
  lead to irregular signs of the coefficients also in higher orders.},
becomes feasible. Only in this case where all coefficients are
positive the corresponding singularity lies on the real axis.
Furthermore, this gives an estimate for the critical temperature where
a possible CEP in the phase diagram might be located. Once this
critical temperature has been obtained the corresponding critical
chemical potential follows immediately via $\mu_c = r(T_c)$.

At higher temperatures the sign of the Taylor coefficients has no
generic structure and the singularity that limits the radius of
convergence of the Taylor series will move in the complex
$\mu$-plane. In this case the convergence radius provides an upper
  bound since the crossover is related to the real part of the
  singularity.

For temperatures below the CEP and larger chemical potentials a
first-order phase transition emerges. As a consequence, two
degenerate minima of the grand potential exist at criticality. Since
the Taylor expansion method involves only local information in the
vicinity of one minimum of the effective potential, it is, in general,
not possible to resolve a first-order phase transition with this
technique. However, it still should be possible to give an upper bound
for the location of the phase boundary.

\section{A model analysis}
\label{sec:model}

In this section we apply the Taylor expansion method to an effective
$N_f = 2+1$ flavor Polyakov-quark-meson model. This model
exhibits a critical endpoint in the $(T,\mu)$ phase diagram and
enables us to inspect the convergence properties of the Taylor series
for \mbox{first-,} second-order and crossover phase boundaries. For
the first time, the Taylor expansion coefficients are calculated to
very high orders, so that the convergence of the series can be probed
reliably. Furthermore, the results are confronted with a full model
solution.

\subsection{The Polyakov-quark-meson  model}

The three flavor Polyakov--quark-meson model~\cite{Schaefer:2009ui} is
a straightforward extension of the two flavor model
\cite{Schaefer:2007pw} and serves as an effective model for
strongly-interacting matter which incorporates a chiral phase
transition, signaled, for example, by a peak in the chiral
susceptibility. Through the coupling to the Polyakov loop, which is an
order parameter for deconfinement in the infinite quark mass limit, it
is also sensitive to deconfining aspects of the QCD phase transition.
The thermodynamics of the PQM model is, for vanishing chemical
potential evaluated in a mean-field approximation, in good agreement
with recent $2+1$ flavor lattice data~\cite{Schaefer:2008ax,
  Schaefer:2009st,*Wambach:2009ee, Schaefer:2009ui}. In contrast to lattice
simulations, the model can be applied directly at finite chemical
potential.

The PQM model Lagrangian reads
\begin{equation}
  \label{eq:lpqm}
  \LPQM = \bar{q}\left(i \Dslash - h \phi_5 \right) q +
  \mathcal{L}_m  
  -\mathcal{U} (\Phi, \Phibar)\ ,
\end{equation}
where $q = (u,d,s)$ denotes the three quark fields. The interaction
between the quarks and the mesons is implemented by a flavor-blind
Yukawa coupling $h$. The meson matrix $\phi_5 =
\sum_{a=0}^8(\lambda_a/2)\left( \sigma_a + i \gamma_5 \pi_a\right)$
consists of nine scalar $\sigma_a$ and nine pseudo-scalar $\pi_a$
meson fields. $\lambda_a$ are the usual Gell-Mann matrices. The
covariant derivative $\Dslash = \dslash - i \gamma_0 A_0$ couples the
Polyakov-loop and its conjugate to the fermionic degrees of freedom
where a gauge coupling has been absorbed in the gauge fields. For
details see \cite{Schaefer:2007pw}.

The purely mesonic contribution is given by
\begin{eqnarray}
\label{eq:mesonL}
  \mathcal{L}_m &=& \Tr \left( \partial_\mu \phi^\dagger \partial^\mu
    \phi \right)
  - m^2 \Tr ( \phi^\dagger \phi) -\lambda_1 \left[\Tr (\phi^\dagger
    \phi)\right]^2 \nonumber \\
  &&  - \lambda_2 \Tr\left(\phi^\dagger \phi\right)^2
  +c   \left(\det (\phi) + \det (\phi^\dagger) \right)\nonumber \\
  && + \Tr\left[H(\phi+\phi^\dagger)\right]\ ,
\end{eqnarray}
with $\phi\equiv \sum_a(\lambda_a/2) \left(\sigma_a + i \pi_a\right)$.
Chiral symmetry is explicitly broken by the last term in
\Eq{eq:mesonL} and the $\ua$-symmetry by the 't Hooft determinant term
with a constant strength $c$. In the $2+1$ flavor scenario only
two order parameters $\vev{\sigma_0}$ and $\vev{\sigma_8}$ emerge in
the singlet-octet basis. They are conveniently rotated to the
non-strange $\vev{\sigma_x}$ and strange $\vev{\sigma_y}$ basis,
see~\cite{Schaefer:2008hk} for details.

The grand potential evaluated in mean-field approximation is a sum
over the mesonic $U$, fermionic $\Omega_{\bar{q}{q}}$ and Polyakov
loop $\mathcal{U}$ contributions:
\begin{equation}
  \label{eq:grandpot}
  \Omega = U \left(\vev{\sigma_{x}},\vev{\sigma_{y}}\right) +
  \Omega_{\bar{q}{q}} \left(\vev{\sigma_{x}},\vev{\sigma_{y}},
    \Phi,\Phibar \right) + 
  \mathcal{U}\left(\Phi,\Phibar\right) \ ,
\end{equation}
where $\Phi$ and $\Phibar$ denotes the real Polyakov loop expectation
values. Note, that at finite chemical potential the $\Phi$ and $\Phibar$ are not linked by complex conjugation and the effective Polyakov loop potential depends on two independent variables.

Explicitly, the quark--anti-quark contribution in presence of the
Polyakov loop reads
\begin{widetext}
  \begin{multline} \label{eq:Omegaqq} \Omega_{\bar qq}(
    \vev{\sigma_{x}} , \vev{\sigma_{y}}, \Phi,\Phibar ) = -2T
    \sum_{f=u,d,s} \int\frac{d^3p}{(2\pi)^3} \left\{\ln \left[1 + 3
        (\Phi + \bar \Phi e^{-(E_{q,f}-\mu)/T})e^{-(E_{q,f}-\mu/T} +
        e^{-3(E_{q,f}-\mu)/T}\right]\right.\\
    + \left. \ln \left[1 + 3 (\bar \Phi + \Phi
        e^{-(E_{q,f}+\mu)/T})e^{-(E_{q,f}+\mu/T} +
        e^{-3(E_{q,f}+\mu)/T}\right] \right\}\
\end{multline}
\end{widetext}
with the flavor-dependent single-particle
energies 
\begin{equation}
  E_{q,f}= \sqrt{k^2 + m_f^2}
\end{equation}
and quark masses
\begin{align}\label{eq:qmasses}
  m_l = h \vev{\sigma_x} /2 &\quad\text{and}\quad m_s = h
  \vev{\sigma_y} /\sqrt{2}
\end{align}
for the light ($l\equiv u,d$) and strange quarks, respectively. In a
mean-field approximation, fluctuations are neglected and a divergent
vacuum contribution to the grand potential has been omitted here which
is irrelevant for our following discussion. See \cite{Skokov:2010sf}
for an investigation of the effect of this vacuum term.

The Yukawa coupling $h$ is fixed such to reproduce a light constituent
quark mass of order $m_{l} \approx 300\MeV$. The parameters of the
mesonic potential $U$ are fitted to well-known pseudo-scalar meson
masses and decay constants~\cite{Schaefer:2008hk}. As
in~\cite{Schaefer:2009ui} we will employ $m_\sigma=600\MeV$.

For the Polyakov-loop part several potentials can be found in the
literature which describe the thermodynamics equally well in the
vicinity of the chiral transition at $\mu=0$, see
e.g.~\cite{Schaefer:2009ui}. Here we employ the logarithmic
version~\cite{Roessner:2006xn}
\begin{eqnarray} 
\label{eq:ulog}
\frac{\mathcal{U}_{\text{log}}}{T^{4}} &=& -\frac{a(T)}{2} \Phibar \Phi\\ 
&&+ b(T) \ln \left[1-6 \Phibar\Phi + 4\left(\Phi^{3}+\Phibar^{3}\right)
  - 3 \left(\Phibar \Phi\right)^{2}\right]\ ,\nonumber 
\end{eqnarray}
with the temperature-dependent coefficients
\begin{equation*}
  a(T) =  a_0 + a_1 \left(\frac{T_0}{T}\right) + a_2
  \left(\frac{T_0}{T}\right)^2\ , \ 
  b(T) = b_3 \left(\frac{T_0}{T}\right)^3
\end{equation*}
and the parameter $T_0=270\MeV$. Note that we ignore any $N_f$- and/or
$\mu$-modifications of this parameter here, i.e., the matter
back-reaction to the gluon sector, in order to simplify our
discussion, see \cite{Schaefer:2007pw, Schaefer:2009ui, Herbst:2010rf}
for more details.

Finally, the temperature and $\mu$-dependent order parameters are
determined by the solution of the gap equations which minimize the
grand potential
\begin{equation}
  \label{eq:pqmeom}
  \left.\frac{ \partial \Omega}{\partial 
      \sigma_x} = \frac{ \partial \Omega}{\partial \sigma_y}  = \frac{
      \partial \Omega}{\partial \Phi}  =\frac{ \partial
      \Omega}{\partial \Phibar} 
  \right|_{min} = 0\ ,
\end{equation}
where {\em min} 
labels the ($T, \mu$)-dependent global minimum of the potential.

\subsection{Higher-orders Taylor coefficients}
\label{sec:ad}
Since the grand potential or the pressure are known, the Taylor
expansion coefficients in \Eq{eq:pcoeff} can be obtained by
calculating the corresponding derivatives of the grand potential
evaluated at the physical potential minimum. This tedious task is
further hindered by the implicit $\mu$-dependence of the order
parameters, introduced by \Eq{eq:pqmeom}. The explicit
$\mu$-dependence is solely given by the fermionic part of the
potential $\Omega_{\bar q q}$ whereof the $\mu$-derivatives can, in
principle, be obtained analytically. However, a fully analytical
approach is infeasible since the implicit dependences can only be
inverted numerically. Standard numerical derivative techniques such as
divided differences may be used for this task at least for the first
few derivatives, but definitely fail for higher orders due to the
increasing rounding and truncation errors.

In order to proceed a novel derivative technique, based on
\emph{algorithmic differentiation} (AD), was developed that properly
allows for the implicit dependencies and further avoids numerical
errors. The AD technique enables the evaluation of derivatives to high
orders even in the case of only numerically solvable implicit
dependences, such as in \Eq{eq:pqmeom}. Furthermore, the derivatives
can be obtained with extremely high numerical precision, essentially
limited only by machine precision. Details of this method can be found
in Ref.~\cite{Wagner:2009pm}.

Of course, this AD approach is not limited to simple Taylor series and
can also be applied to more involved Taylor expansions of physical
quantities which also require temperature-derivatives. An example
would be the Taylor expansion of the trace anomaly $(\epsilon -3p)$.

\begin{figure}[tb]
  \centering
  \includegraphics[width=\onefig]{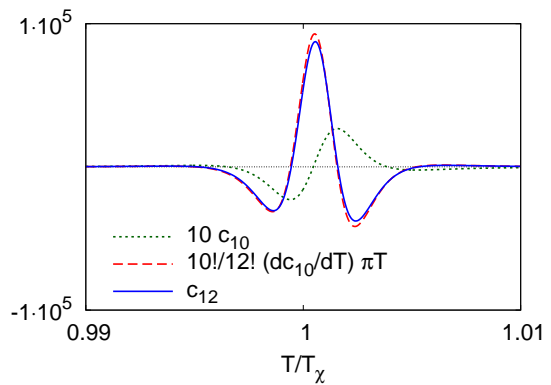}
  \caption{The Taylor coefficient $c_{12}$ and the (rescaled)
    temperature derivative of the coefficient $c_{10}$ in the vicinity
    of the critical temperature $T_\chi$. For comparison the
    (rescaled) Taylor coefficient $c_{10}$ is also shown.}
\label{fig:coeff}
\end{figure}

The Taylor coefficients $c_n$ for the pressure, \Eq{eq:pcoeff} have
been calculated with this novel AD technique for the three flavor PQM
model and can be found, up to the $22^{\textrm{nd}}$ order,
in~\cite{Schaefer:2009st, *Wambach:2009ee}. All coefficients for $n>4$ show rapid
oscillations near the transition temperature $T_\chi$ at $\mu=0$. The
coefficients become small outside a 5\% temperature interval around
the $\mu=0$ transition temperature $T_\chi$, i.e. $0.95 < T/T_\chi <
1.05$. Within this temperature interval the amplitudes of the
oscillations increase with increasing order and hence higher orders
are important even for small expansion parameters, i.e.~$(\mu/T) < 1$.
As an example the coefficient $c_{12}$ and the properly rescaled
temperature derivative of the coefficient $c_{10}$ are shown in
\Fig{fig:coeff}. As already discussed in Sec.~\ref{sec:locating} it
obeys the same singular behavior as $c_{12}$, hence the structure of
$c_{12}$ can be estimated by the temperature-derivative of $c_{10}$.
Interestingly, even the magnitude of the coefficient $c_{12}$ is well
estimated by dimensional analysis and compensating the different
factorials.
Some convergence analysis as well as detailed extra\-polations of the
pressure and quark number susceptibility Taylor series to finite $\mu$
with these coefficients can be found in~\cite{Schaefer:2009st,
  *Wambach:2009ee}.

\subsection{Determining the phase boundary}

We start with a calculation of the $(T,\mu)$ phase diagram of the PQM
model without referring to a Taylor expansion. At vanishing chemical
potential the chiral crossover, signaled here by a peak in the chiral
susceptibility occurs at $T_\chi\sim206$ MeV. At this temperature also
the Polyakov loop susceptibility has its maximum. For non-vanishing
chemical potentials the position of maxima in both susceptibilities
differ. The peak in the Polyakov loop susceptibility occurs at higher
temperatures than that in the chiral susceptibility. This scenario
changes, however, when fluctuations are taken into account, e.g., if
$\mu$-corrections of the $T_0$ parameter in the Polyakov loop
potential are considered, see \cite{Schaefer:2007pw, Herbst:2010rf}.
For small temperatures a first-order chiral phase transition is found
which terminates at larger temperature in a CEP. In mean-field
approximation the CEP is located at $(T_c,\mu_c) \sim (185, 167)\MeV$
which corresponds to $T_c\sim0.9\ T_\chi$ and $\mu_c/T_c \sim 0.9$.
Approximations beyond mean-field push the location of the CEP off the
$T$-axis. For a mean-field and renormalization group comparison in a
quark-meson model, see e.g., \cite{Schaefer:2006ds}. In the following
we compare several methods for the determination of the phase boundary
with the Taylor coefficients.

\subsubsection{Taylor expansion}
\label{sec:phaseboundary}

\begin{figure}[tb]
\centering
\includegraphics[width=\onefig]{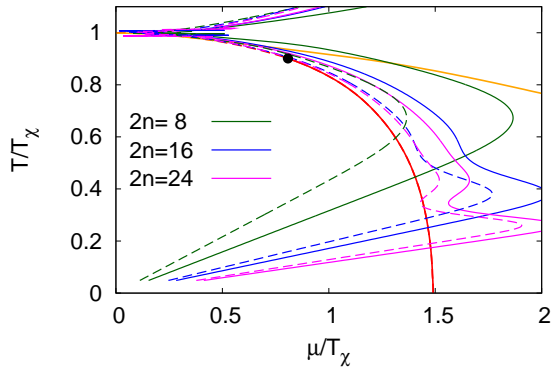}
\caption{Estimates for the radius of convergence obtained from
  $r_{2n}$ (solid lines) and $r_{2n-2}^\chi$ (dashed lines) for
  different orders ($2n=8,16,24$) of the Taylor expansion. Also shown
  are the phase boundaries for the chiral (red line; dashed:
  crossover, solid: first order) and a deconfinement line identified
  by the peak position of the Polyakov loop susceptibility (yellow
  line). The black dot indicates the location of the CEP.}
\label{fig:convrad}
\end{figure}


In \Fig{fig:convrad} we show the results for two estimators for the
radius of convergence, obtained from the expansion coefficients of the
pressure, $r_{2n}$ (solid lines), and of the quark number
susceptibility, $r^\chi_{2n-2}$ (dashed lines), for three different
truncation orders ($2n=8,16,24$) in the ($T,\mu$) phase diagram. The
phase boundary for the chiral transition and a deconfinement line,
corresponding to a maximum in the Polyakov loop susceptibility, are
also displayed. Note that the difference in the radii of convergence,
estimated from the $r_{2n}$ and $r^\chi_{2n-2}$, respectively, is only
caused by the prefactor in \Eq{eq:convradchi} since the same Taylor
coefficients contribute to both estimators.

Near $T_\chi$ the sign of the expansion coefficients oscillates,
indicating that the break down of the series expansion is caused by a
singularity in the complex plane away from the real axis. The
oscillations in the Taylor coefficients entail oscillations in the
radii of convergence at small chemical potentials which are hard to
see in \Fig{fig:convrad}. Hence, for small chemical potentials the
radius of convergence is not suitable to estimate the phase boundary.
For larger values of the chemical potential and for temperatures below
$T_\chi$ estimates for the radius of convergence approach the phase
boundary from above with increasing truncation order $n$. The observed agreement suggests that the singularity limiting the Taylor expansion is close to the real $\mu$-axis and the small imaginary part is negligible when compared to the error due to the finite number of coefficients available for a fast crossover. In
 particular, this is still valid for the region in the vicinity of the
CEP. We note that for $n\gsim 10$ corrections to the asymptotic value
$r$ are well described in terms of ${\cal O}(1/n)$ corrections. In
fact, with an ansatz $r (1+A/n)$ for the estimators of the radius of
convergence $r_{2n}$, the leading ${\cal O}(1/n)$ corrections can be
eliminate by using subsequent estimates.
This leads to estimates for $r$ obtained from the pressure and quark
number susceptibility series that agree within a few percent and reach
the asymptotic value within 15\% for $2n>14$. The coefficient $A$ for
the correction term, obtained in this way, is about a factor 5 larger
in the pressure series than in the quark number susceptibility series.
Qualitatively, this different asymptotic behavior of both series can
be understood in terms of the structure of the singular contribution
to the grand potential that limits the radius of convergence. We
discuss this in more detail in the Appendix \ref{sec:crit}.

At low temperatures the estimates for the radius of convergence
increase with increasing order $n$ of the Taylor expansion and allow
also ratios $\mu/T > 1$. This is in agreement with the small magnitude
of the higher order coefficients in this temperature region. 
In this region the results are in agreement with results from the resonance gas (cf.~~\cite{Karsch:2003vd, *Karsch:2003zq}), i.e. the estimates of the convergence radius of the pressure are given by $r_{2n}^{\text{HRG}}(T)=1/3 \sqrt{(1+2n)(2+2n)}$. 

As already mentioned above, it is not possible to resolve a first-order
transition with the Taylor expansion. Therefore, $r_{2n}$ as well as
$r^{\chi}_{2n}$ yield estimates for the radius of convergence that are
larger than the true location of the first-order transition line and
signal a misleading convergence of the Taylor expansion. In this
region the Taylor expansion might provide stable and continuous
results but one cannot extract the precise position of the boundary
for a first-order phase transition.

\subsubsection{Pad\'e approximation}

\begin{figure}[tb]
\centering
\includegraphics[width=\onefig]{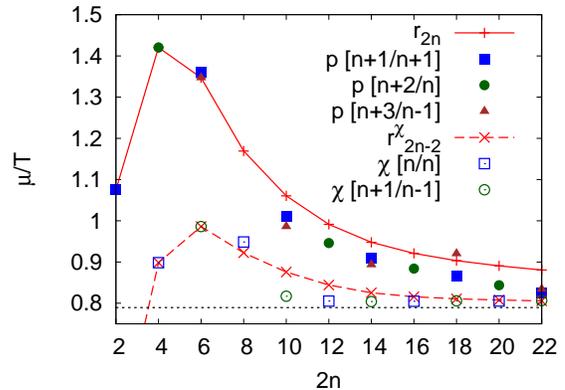}
\caption{Estimate of the phase boundary at
  $T=190\MeV \sim0.92\ T_\chi$ with $r_{2n}$ and $r^\chi_{2n}$ and
  poles in the \pade{} approximation of the pressure and quark number
  susceptibility as a function of the order of the Taylor expansion
  $2n$. The used highest coefficient is $c_{2n+2}$. The dotted
  horizontal line indicates the phase boundary calculated directly at
  finite $\mu$. Note also that an estimate is possible only for even
  \pade{} approximations. }
\label{fig:pbpade}
\end{figure}

The \pade{} approximation could improve the convergence behavior and
therefore provide a better estimate of the phase boundary since the
\pade{} approximant employs all Taylor coefficients in contrast to the
previous analysis where only two subsequent Taylor coefficients enter.
The pole in the $[N/2]$ \pade{} approximation yields the corresponding
radius of convergence as described in Sec.~\ref{sec:pade_approx}. For a
general $[N/M]$ \pade{} approximation we use the first pole at real
and positive $\mu$ in order to estimate the phase boundary. Since all
Taylor coefficients with their corresponding error enter in the
\pade{} approximant the error propagation is more involved here in
contrast to the previous discussion where only two error sources
enter. In this context, the application of the \pade{} approximation
in lattice simulations is much more involved. However, in the model
analysis the coefficients, obtained with the AD technique, exhibit
extremely small numerical errors and result in a stable and reliable
\pade{} approximation. 

Apart from the diagonal scheme $[N/N]$ we also use a non-diagonal
$[N+2/N]$ approximation for the susceptibility and a $[N+2/N-2]$,
respectively $[N+2/N]$ approximation for the pressure and determine
always the first pole at real and positive $\mu$.

In~\Fig{fig:pbpade} we show for a fixed temperature $T=0.92\; T_\chi$
the estimates of the phase boundary extracted from \pade{}
approximations for the pressure ($p$) and the quark number
susceptibility ($\chi$) for different truncation orders $n$. The
chosen temperature is slightly above the temperature of the CEP in the
phase diagram. In addition, the estimates for the radii of convergence
($r_{2n}$ and $r^\chi_{2n}$) and the chiral crossover chemical potential (dashed
horizontal line) are also shown. For small truncation orders
$2n \leq 8$ the $[N/2]$ \pade{} approximations coincide
with the radii of convergence as expected. For $2n >8$ we
observe that the \pade{} approximation converges faster for the
pressure as well as for the susceptibility compared to the radii of
convergence obtained without the \pade{} approximation.

We conclude that the \pade{} approximation converges faster at larger
truncation orders, in particular for the quark number susceptibility.
The estimate of the phase boundary from the \pade{} approximation
becomes comparable to the one obtained from estimators for the radius
of convergence determined directly from ratios of Taylor expansion
coefficients only at significantly larger truncation order. For
example, in \Fig{fig:pbpade} the distance of the \pade{} estimate to
the horizontal line at $2n=12$ is achieved with the Taylor expansion
only for $2n\geq20$. However, the lower truncation order in the
\pade{} approximation induces a more involved error propagation. This
is no problem with the AD-techniques but might hamper lattice
simulations.

\paragraph{First-order transition}

As already mentioned, the Taylor expansion around $\mu=0$ fails if a
first-order phase transition emerges since the jump to a new global
minimum of the effective potential cannot be resolved. However, 
it might still be possible to enter the metastable phase with the
Taylor expansion method. In this case the radius of convergence just
reflects the fact that this expansion is still valid in the metastable
phase.  However, a precise determination of the first-order transition is
not possible. This conceptual drawback is not modified when
the \pade{} approximation is employed.

\begin{figure}[tb]
\centering
\includegraphics[width=\onefig]{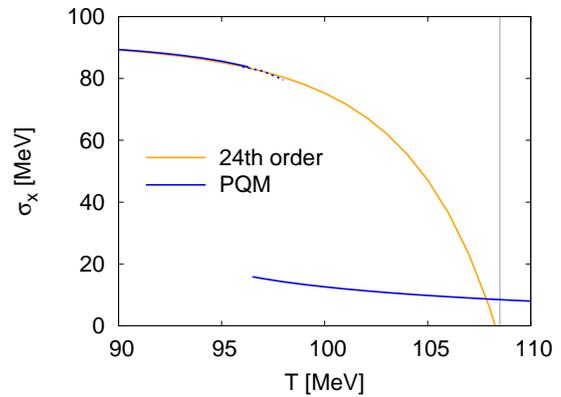}
\caption{\label{fig:sxtaylor}Taylor expansion of the (non-strange)
  order parameter $\sigma_x$ compared to the direct evaluation at
  finite $\mu$ for $\mu/T=3$. The dotted lines indicate the beginning
  of the metastable phase. The vertical line indicates the estimate
for the radius of convergence, $r_{24}$.}
\end{figure} 

As an example we show in \Fig{fig:sxtaylor} for $\mu/T=3$ the chiral
order parameter, the non-strange condensate $\sigma_x$, as a function
of the temperature in the vicinity of a first-order transition. Here
$\sigma_x$ has been obtained from a Taylor series truncated at
24$^{th}$ order. The lines, labeled with PQM, represent the full model
solution of the gap equation \Eq{eq:pqmeom} where a first-order
transition can be resolved. Both solutions agree very well in the low
temperature phase until $T \sim 96$ MeV. The dotted line indicates the
begin of the metastable phase. One sees that it is still possible to
enter this metastable phase with the Taylor expansion method. The
vertical line in the figure marks the location of the estimate for the
radius of convergence obtained for the chosen $n=24$ truncation order.
Furthermore, this example also demonstrates that the Taylor expansion
method is applicable also for $\mu/T > 1$ as long as one stays within
the convergence regime.

\paragraph{Critical endpoint}

Directly at the critical endpoint a second-order phase transition
emerge and no discontinuity in the order parameter appears. Hence, the
radius of convergence might be well-suited to estimate the location of
the critical endpoint in the phase diagram once the critical
temperature is known. The corresponding critical chemical potential
can again be obtained via
\begin{equation}
\label{eq:muctc}
\mu_c = r(T_c) = \lim_{n \rightarrow \infty} r_n(T_c) \ .
\end{equation}
 
From the discussion given so far and from Fig.~\ref{fig:pbpade} one
may conclude that at fixed temperature the radius of convergence can
be estimated with an accuracy of about 15\%-20\% from a series of the
order ${\cal O}(\mu^{12})$. This is clearly an acceptable uncertainty
in view of the difficulties one has to face in current QCD
calculations with non-zero chemical potential.

However, the determination of $T_c$ is still a non-trivial task. A
criterion is needed, that allows to estimate at a given order of the
expansion the temperature regime in which all expansion coefficients
may stay positive as discussed in Sec.~\ref{sec:locating}. As an
example we show the zeros of the Taylor series of the pressure in the complex
$\mu/T$-plane around a temperature point where a coefficient changes
its sign in Appendix~\ref{sec:approots}.

\begin{figure}[t]
\centering
\includegraphics[width=\onefig]{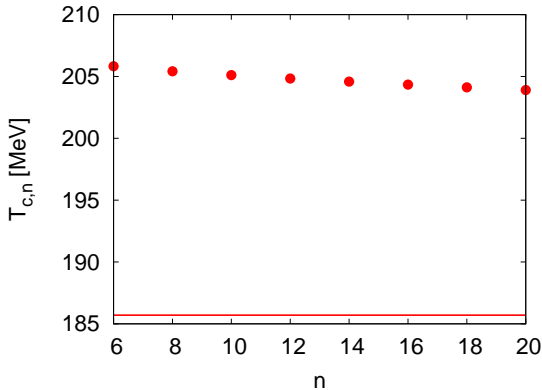}
\caption{Temperature $T_{c,n}$ at which the Taylor coefficients $c_n$
  becomes negative for different truncation orders $n$. The solid
  horizontal line denotes the critical temperature of the CEP.}
\label{fig:taylorzero}
\end{figure}

In Fig. \ref{fig:taylorzero} we plot the temperature $T_{c,n}$ defined
as the temperature where the Taylor coefficient $c_n$ becomes negative
for different truncation orders. This yields an upper bound for the
critical temperature. Of course, in the limit $n\rightarrow \infty$
the temperature series of these approximations should approach the
critical temperature of the CEP, i.e.,
\begin{equation}
T_c = \lim_{n\rightarrow \infty} T_{c,n}\ .
\end{equation}
However, at least with our model analysis, only a very slow
convergence of the estimators for the critical temperature at the CEP
is visible. Even at high truncation order of 20$^{\text{th}}$ there
is still a temperature difference of the order of $20$ MeV which is why we do not
reconsider \Eq{eq:muctc}.

In addition, this convergence behavior might be improved by applying
the \pade{} approximation but the identification of a proper
convergence criterion is more involved as discussed in
Appendix~\ref{sec:approots}.

%
%
\section{Summary and conclusions}
\label{sec:summary}

The convergence properties of the Taylor expansion method which is
used to extend lattice QCD calculations to non-zero chemical potential
have been investigated. The method has been applied to an effective
$N_f=2+1$ flavor Polyakov--quark-meson model and the results obtained
with the Taylor expansion have been compared to a full model
mean-field solution. By means of a novel algorithmic differentiation
technique higher-order Taylor coefficients could be calculated for the
first time. As a function of the temperature the coefficients start to
oscillate in the vicinity of the phase transition region. The number
of roots and magnitude of the oscillations increase with the
coefficient order. Two definitions of the radius of convergence, one
through the pressure series coefficients and the other one via the
corresponding susceptibility coefficients have been discussed.
The convergence of the estimators for the radius of convergence of the susceptibility series is  faster
than the corresponding one for the pressure series.

With the \pade{}
approximation the convergence rate could be further improved. By
investigating the pole structure of the \pade{} approximant a
definition for the radius of convergence could be extracted and a
better estimate for the phase boundary could be achieved.

While the convergence radius yields the precise location of the
  phase boundary only for a second-order transition, i.e., at the
  critical endpoint, we observed a good agreement of the convergence
  radius with the full model solution also for temperatures above
  $T_c$. In view of the error of convergence radius due to the finite
  number of coefficients the convergence radius can provide a
  valuable estimate also for the location of a rapid crossover
  where the limiting singularity is close to the real axis in the
  complex $\mu$-plane. 
 
The singularity lies on the real axis only in the case where all coefficients are positive. This criterion also provides an estimate for the critical temperature $T_{c,n}$. In our study this estimate depends only weakly on the truncation order. But even
  for truncation orders above $20^{\text{th}}$ there is still a
  significant gap to the full model solution and a precise
  determination of the location of the CEP remains open. Note that the
  error in estimating $T_c$ also affects the estimation of $\mu_c$.


Although, the situation may be different in model calculations and in
QCD, our study suggests that for a meaningful application of the
Taylor method to lattice QCD higher order Taylor coefficients are
definitely necessary in order to control systematically truncation
errors in the series expansion and to estimate reliably the location of a
possible critical endpoint in the QCD phase diagram. At present the
available number of coefficients extracted in lattice QCD calculations
is not sufficient to locate a critical endpoint in the QCD phase
diagram with the Taylor expansion method. However, if the critical
endpoint is located at smaller temperatures a broader temperature
interval is expected to emerge in which the Taylor coefficients the
Taylor coefficients will oscillate. The estimators for the critical
temperatures $T_{c,n}$ are then expected to show a stronger
temperature dependence and they should decrease faster. However, in
this case also the CEP may be located at large values of $\mu/T$ and
higher orders in the expansion may be needed due to this.

Furthermore, when going beyond mean-field approximations in our model
calculation, quantum and thermal fluctuations in the hadronic phase
will surely modify the oscillations of the Taylor coefficients. This
too may improve the estimate of the critical temperature with the
Taylor expansion method. However, also in this case the location of a
possible CEP is shifted towards the $\mu$-axis \cite{Nakano:2009ps}
and more coefficients are necessary for an adequate description. To
explore this quantitatively we plan to extend the present study beyond
mean-field as well as tightening the AD techniques with functional
renormalization group methods.

With the Taylor expansion method it is in general not possible to
describe a first-order transition completely including the precise
determination of the critical temperature since only information about
one potential minimum are available. One may, however, establish upper
bounds for the location of the phase boundary.

\section*{Acknowledgment}
\begin{figure*}[Tt]
\centering
\subfigure[ $\,c_{16} > 0$]{
\includegraphics[width=\threefigs]{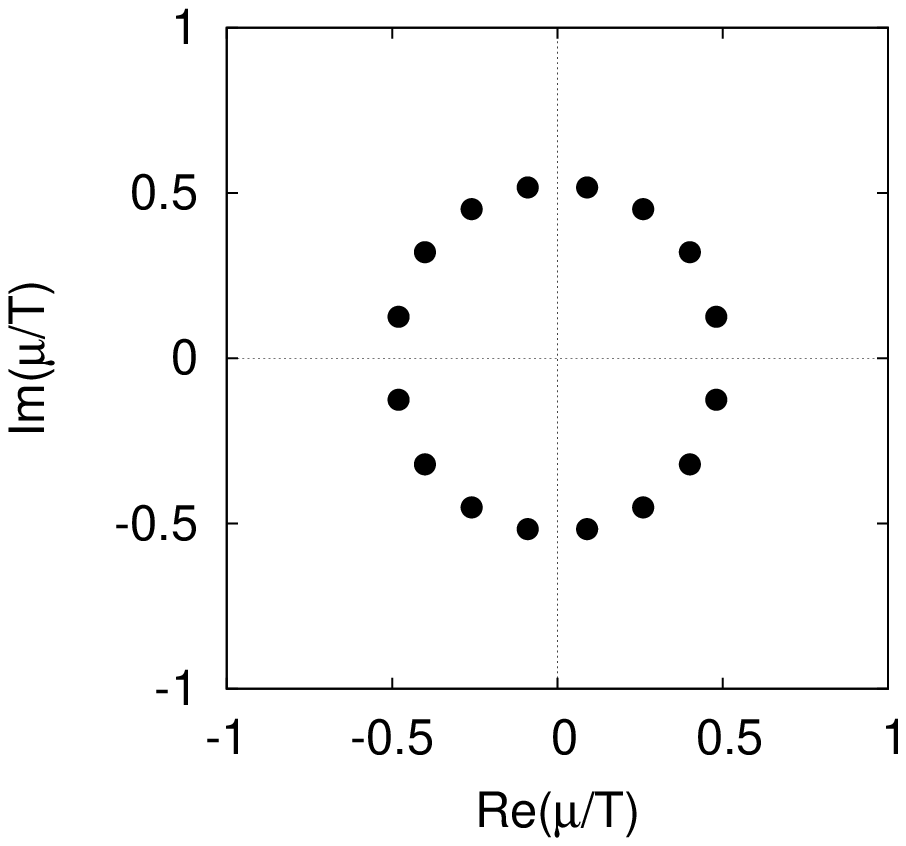}
}
\subfigure[ $\,c_{16} = 0$]{
\includegraphics[width=\threefigs]{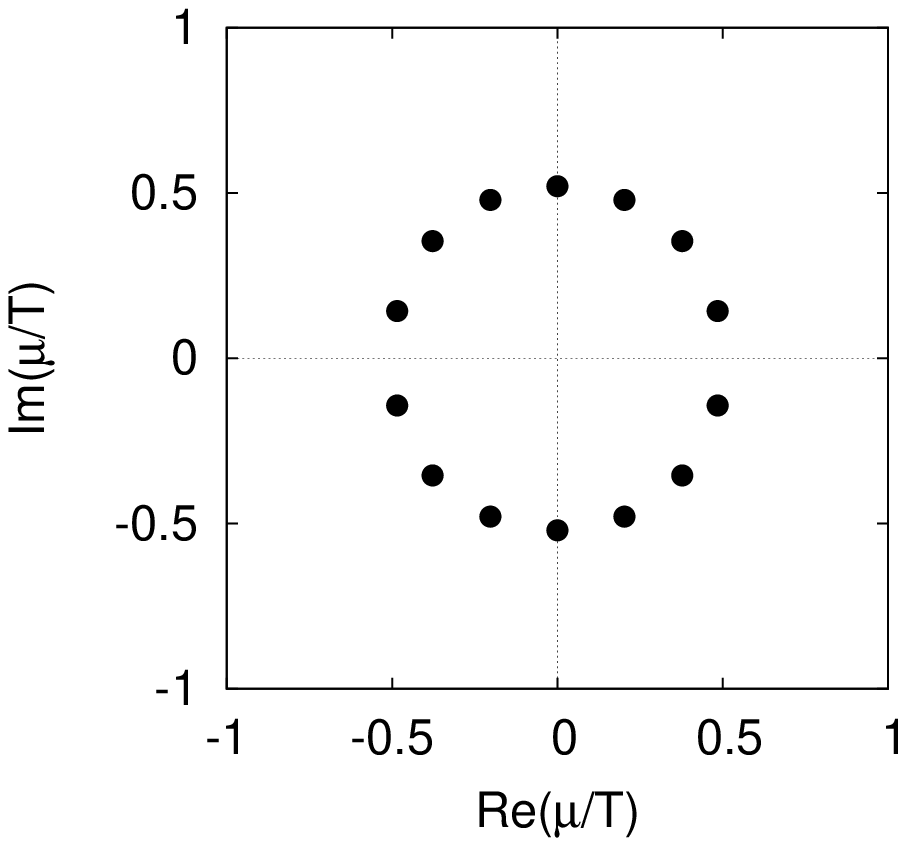}
}
\subfigure[$\,c_{16} < 0$]{
\includegraphics[width=\threefigs]{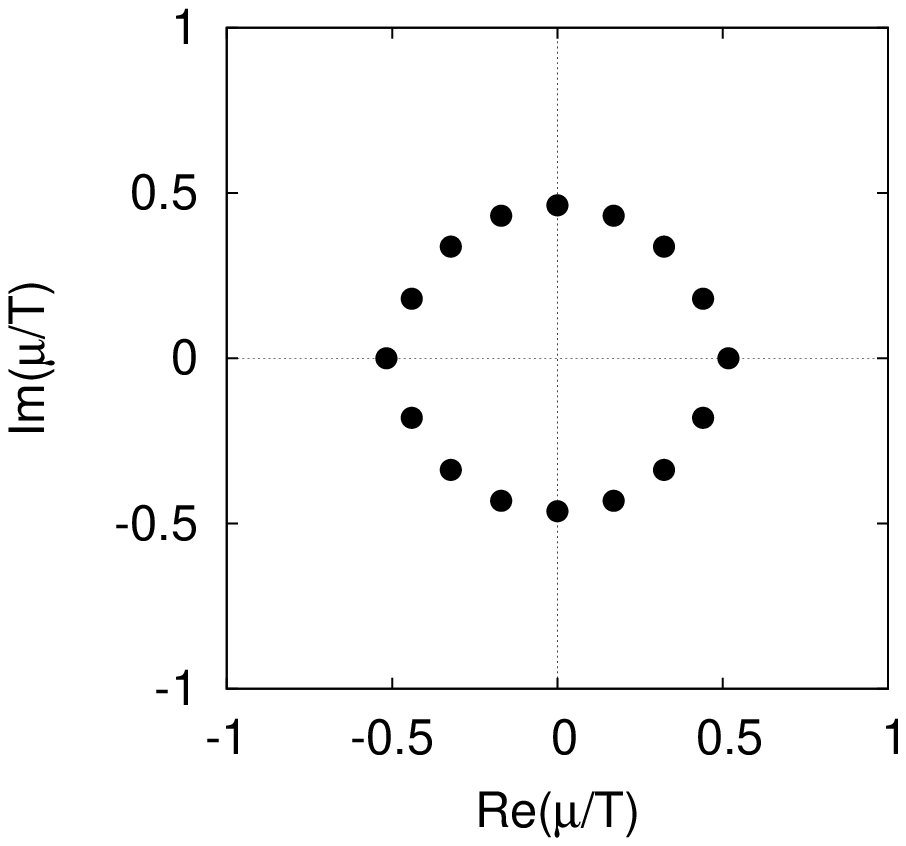}
}
\caption{\label{fig:ylzero} Root distribution in the complex
  $\mu/T$-plane of a $16^{\text{th}}$ order Taylor series of the
  pressure. The corresponding temperatures in the three panels are
  chosen such that the highest Taylor coefficient $c_{16}$ changes
  sign while all the remaining Taylor coefficients are still positive.
}
\end{figure*}

We gratefully acknowledge very useful discussions with Christian
Schmidt. The work of FK has been supported in part by contracts
DE-AC02-98CH10886 with the U.S. Department of Energy and the BMBF
under grant 06BI401. MW acknowledges support by the Alliance Program
of the Helmholtz Association (HA216/ EMMI) and BMBF grant 06DA9047I.
The work of MW and JW were supported in part by the Helmholtz
International Center for FAIR.

\appendix

\section{Asymptotic behavior of Taylor expansion coefficients}
\label{sec:crit}

The QCD critical point is expected to belong to the universality class
of the three dimensional Ising model. The critical behavior in the
vicinity of this point is controlled by two couplings which
characterize fluctuations in thermal and magnetic (symmetry breaking)
directions of the effective Ising Hamiltonian. The chemical potential
will mix with both couplings \cite{Karsch:2001nf}. When approaching
the critical point in the $(T,\mu)$-plane the dominant singular
contributions to, e.g., the quark number susceptibility arise from the
magnetic coupling. When approaching the critical endpoint in the QCD
phase diagram at fixed temperature, the quark number susceptibility
thus is expected to scale like \cite{Karsch:2001nf, Hatta:2002sj,
  Schaefer:2006ds}
\begin{equation}
\chi_q (T,\mu)/T^2 \sim (1-\mu/\mu_c)^{1/\delta -1} = 
(1-\mu/\mu_c)^{\gamma/\beta\delta} 
\label{crit}
\end{equation}
which arises from a singular term in the grand potential, i.e., the
pressure which has the form,
\begin{equation}
\left( p/T^4\right)_\text{sing} \sim (1- \mu/\mu_c)^{1+1/\delta} \; .
\label{pcrit}
\end{equation}
Assuming that these singular terms give indeed the dominant
contribution to the Taylor expansion coefficients at high order, we
obtain for the asymptotic behavior of the estimators for the radius of
convergence ($r\equiv \mu_c$) of the pressure and quark number
susceptibility series, respectively,
\begin{eqnarray}
r_{2n} &\sim& \mu_c \left( 1 + \frac{2\delta+1}{2n\delta} \right) \nonumber \\
r_{2n}^{\chi} &\sim& \mu_c \left( 1 + \frac{1}{2n\delta} \right) \; .
\label{asymptotic}
\end{eqnarray}
We thus expect that in comparison to the pressure series, corrections
to the radius of convergence are smaller in the quark number
susceptibility series. In mean-field approximation, which is used in
our current study, the relative factor is $2\delta+1 =7$ and it is
about $10.6$ in the three dimensional Ising model. Indeed, this seems
to be realized with the series discussed in this study, although the
approach to the asymptotic value is slow.
\vspace{2ex}

\section{Roots in the complex $\mu/T$-plane}
\label{sec:approots}

As mentioned in Sec.~\ref{sec:locating} a singularity in the complex
$\mu/T$-plane limits the radius of convergence of a Taylor expansion.
In Fig.~\ref{fig:ylzero} an example of the root distribution in the
complex $\mu/T$-plane of a $16^{\text{th}}$ order Taylor series of the
pressure is given. The temperatures in the three panels are chosen
such that the highest Taylor coefficient $c_{16}$ changes sign while
all the remaining coefficients stay positive. In the left panel
$c_{16}$ is still positive. As expected all 16 roots are complex and
distributed on a circle whose radius corresponds to the radius of
convergence $r_{16}$. For $c_{16}= 0$ (middle panel) the Taylor
polynomial is of degree 14 and consequently only 14 roots emerge.
Approaching $c_{16}=0$ from positive values the two root pairs which
are each closest to the imaginary axis merge to each one purely
imaginary root and the other root vanishes at imaginary plus or minus
infinity. When the temperature is further increased, i.e., $c_{16}$
becomes negative (right panel), two roots reenter the complex
$\mu/T$-plane on the real $\mu$-axis from plus or minus infinity and
rapidly approach the radius of convergence.

A similar analysis could also be performed with the \pade{}
approximation which shows, in general, a more rapid convergence
behavior. However, the investigation for this case is more involved
since two polynomials occur and possible root cancellations can emerge
which are hard to distinguish numerically from real roots. In
addition, deeper problems may further arise since the CEP is related
to a pole with a non-integer critical exponent which is generally
difficult to describe properly within a \pade{} approximation.


%

\end{document}